# Electrical switching of the perpendicular Néel order in a collinear antiferromagnet

Wenqing He[1,2,†], Tianyi Zhang[1,†], Yongjian Zhou[3,†], Caihua Wan*[1,4], Hao Wu[4], Baoshan Cui[4], Jihao Xia[1], Ran Zhang[1], Tengyu Guo[4], Peng Chen[1], Mingkun Zhao[1], Leina Jiang[1], Alexander Grutter[5], Purnima P. Balakrishnan[5], Andrew J. Caruana[6], Christy J. Kinane[6], Sean Langridge[6], Guoqiang Yu[1,4], Cheng Song[3], Xiufeng Han*[1,2,4]

[1]Beijing National Laboratory for Condensed Matter Physics, Institute of Physics, Chinese Academy of Sciences, Beijing 100190, China

[2]Center of Materials Science and Optoelectronics Engineering, University of Chinese Academy of Sciences, Beijing 100049, China

[3] Key Laboratory of Advanced Materials (MOE), School of Materials Science and Engineering, Tsinghua University, Beijing 100084, P. R. China

[4]Songshan Lake Materials Laboratory, Dongguan, Guangdong 523808, China

[5]NIST Center for Neutron Research, National Institute of Standards and Technology, Gaithersburg, Maryland 20899-6102, United States;

[6]ISIS Facility, STFC Rutherford Appleton Laboratory, Harwell Science and Innovation Campus, Oxon, OX11 0QX, United Kingdom.

**Electrical manipulation of magnetic order by current-induced spin torques lays the foundation for spintronics. One promising approach is encoding information in the Néel vector of antiferromagnetic (AFM) materials, particularly to collinear antiferromagnets with the perpendicular magnetic anisotropy (PMA), as the negligible stray fields and terahertz spin dynamics can enable memory devices with higher integration density and ultrafast speed. Here we demonstrate that the Néel order information in a prototypical collinear AFM insulator with PMA, $Cr_2O_3$, can be reliably readout via the anomalous Hall effect and efficiently switched by the spin-orbit torque (SOT) effect with a low current density of $5.8×10^6$ A/cm$^2$. Moreover, using $Cr_2O_3$ as a mediator, we electrically switch the magnetization of a $Y_3Fe_5O_{12}$ film exchange-coupled to the $Cr_2O_3$ layer,**

**unambiguously confirming the Néel order switching of the $Cr_2O_3$ layer. This work provides a significant basis for developing AFM memory devices based on collinear AFM materials with PMA.**

For decades, ferromagnet (FM)-based spintronics have led to the birth of high performance spintronic devices, such as magnetic random-access memory (MRAM).[1-3] The success of MRAM was based on the development of magnetic tunnel junctions (MTJ), whereby the resistance can be reliably read out through the tunneling magnetoresistance effect and whose magnetic configurations can be electrically written with high efficiency via the current-induced spin torques. In particular, spin-orbit torque (SOT) originating from spin-orbit coupling in the heavy metal (HM) layer offers unique advantages in efficiency, speed and flexibility[3]. To further improve the performance of memory devices, antiferromagnetic (AFM) materials have recently attracted significant interest[4-6]. The natural properties of AFM materials – zero stray-field and THz spin dynamics[7-9] – enable the realization of higher density and ultrafast devices, with immense potential for next-generation memory technologies[6,10-16].

AFM materials can be categorized into collinear and noncollinear ones according to their specific AFM textures. Inspiringly, both collinear and noncollinear AFMs are recently predicted to exhibit large tunnelling magnetoresistance (TMR) when they are used in antiferromagnetic tunnel junctions (AFMTJs)[11-13]. For instance, AFMTJs comprising collinear AFM $RuO_2$ have predicted TMR ratios above 500%[13]. Recently, AFMTJs, made of noncollinear $Mn_3Pt$ or $Mn_3Sn$, have experimentally demonstrated a sizeable TMR effect at room temperature[11-12]. However, efficient electrical AFM writing remains a challenge that requires further optimization. Combining AFMTJs with efficient electrical switching is essential for significant progress towards practical AFM spintronics.

Many AFM materials have multiple easy axes along which their Néel order can be aligned, instead of only two that can directly encode the binary states 0 and 1 as in ferromagnets[14-22]. Full switching, which refers to the rotation of a magnetic order parameter (magnetization for FM or Néel order for AFM) by 180°, can directly encode the 0 and 1 binary data straightforwardly as desired. Such 180º switching has several advantages over 90º or other switching strategies for AFM memories [23-24]. Firstly, such a full switching phenomenon is critical for rapid operation speed because it maximizes the readout margin to determine the state of a device, which in turn reduces the time required to do so. Secondly, 180º switching allows a simpler device structure and therefore a denser package, which further increases the storage density relative to 90º switching. Thirdly, the higher magnetic anisotropy barrier separating the two memory states can ensure a higher retention and stronger robustness against external influences. Recent work has investigated full electrical switching of the noncollinear AFM metal $Mn_3Sn$ films in which tensile strain from substrates is applied to realize PMA[24]. Electrical switching of a collinear AFM by 90º has also been reported, though debated in some cases due to possible entanglement with the electromigration and thermoelastic effects [25-26]. For applications it is therefore important to demonstrate full 180º switching of a perpendicular Néel order in a collinear AFM, which is a long-term goal for developing ultra-dense and ultra-fast antiferromagnetic memory with opposite Néel vectors as binary states 0 and 1. However. the current switching mechanisms have still been limited for 90° or 120° switching in AFMs, the unique properties and functionalities of electrical switching of perpendicular Néel order in a collinear AFM have yet to be unveiled.

Though collinear AFMs are simpler in their spin structure, it is difficult to unambiguously demonstrate 180º full switching of a collinear Néel vector. To convincingly demonstrate this switching requires three characteristics: (1) The desired 180º switching cannot be readout by the planar Hall effect anymore which has been extensively exploited for the case of 90º rotation. Therefore, a significant anomalous Hall effect is necessary for 180º switching. (2) The collinear AFM state should preferably have a perpendicular magnetic anisotropy, which helps to distinguish any influences from the thermal and Oersted field effects. (3) The Néel order switching of the collinear AFM should also be clearly distinguishable from the case of switching of uncompensated spins at the interface only. Consequently, unambiguous demonstration of the collinear Néel order switching in experiment is challenging.

$Cr_2O_3$ is a collinear AFM insulator with its easy-axis along the *c* axis of a hexagonal lattice[27] (Fig.1**a**). Because of its intriguing magnetoelectricity, antiferromagnetism and insulating properties[7,27-29], $Cr_2O_3$ is an important material for AFM magnonics[7] and multiferroics[28,29], attracting broad interests. Here, we report the SOT-induced 180º full switching of the Néel order of the collinear AFM $Cr_2O_3$ with PMA. Its PMA allows us to clearly detect the 180º reversal of its collinear Néel order by an AHE-like phenomenon resulting from interfacial spins that directly couple to the Néel order. Furthermore, not only the Néel order of $Cr_2O_3$ but also the perpendicular magnetization of a $Y_3Fe_5O_{12}$ (YIG) film spaced by $Cr_2O_3$ is reversed via SOT in the YIG/$Cr_2O_3$/Pt control devices, which confirms the 180º switching of the Néel vector throughout the entire $Cr_2O_3$ rather than uncompensated interfacial spins. This work demonstrates the feasibility of developing electrically operational memories based on collinear AFM materials, greatly revealing a route to using collinear and perpendicular spin structures in AFM spintronics.

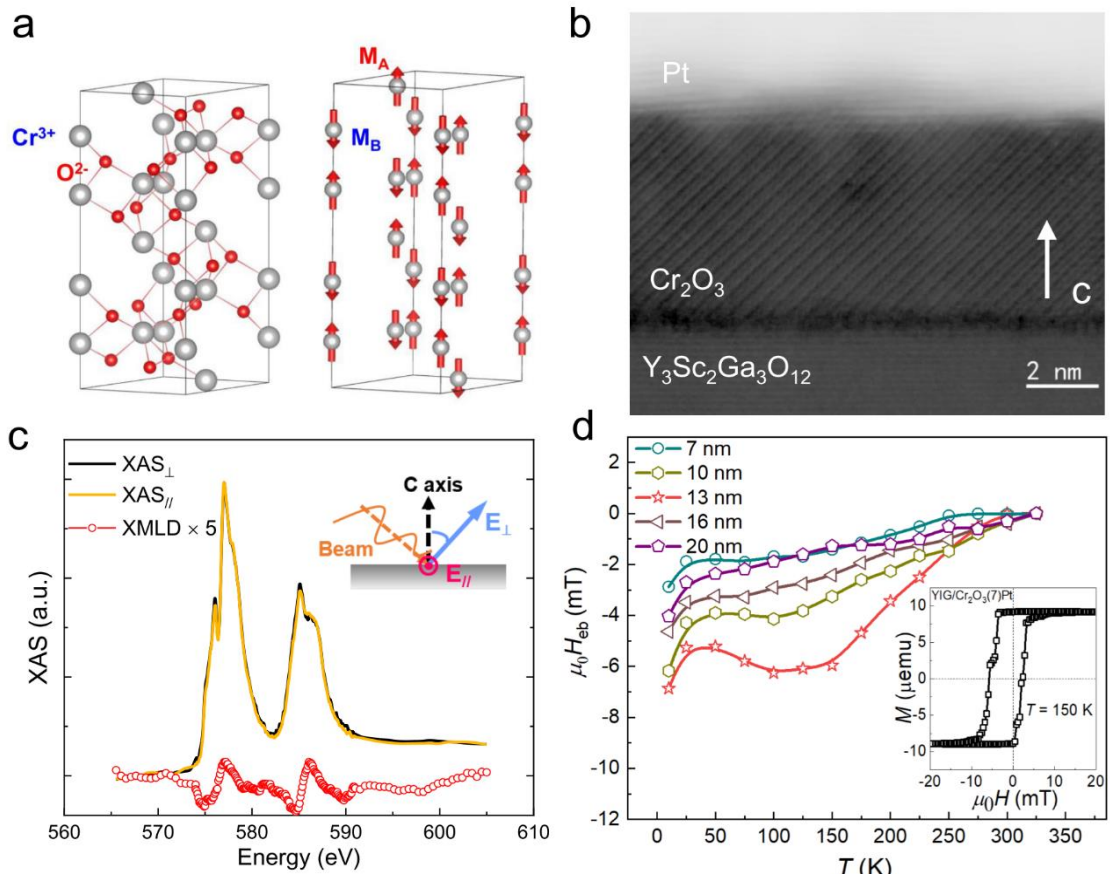

**Figure 1. Atomic and magnetic structures of Cr$_2$O$_3$. a**, Atomic and AFM structures of a Cr$_2$O$_3$ unit cell. **b**, The cross-sectional HAADF image of the YSGG//Cr$_2$O$_3$(7 nm)/Pt(3 nm) heterostructure. **c**, The XAS and XMLD spectra of the Cr$^{3+}$ $L_3$ and $L_2$ edges in YSGG//Cr$_2$O$_3$/Pt at 180 K. The polarization dependence was obtained with more than 98% linearly polarized light with an incident angle of 45° from the c-axis. **d**, The $T$-dependence of the exchange bias field of the samples with different $t_{Cr_2O_3}$. Inset shows a typical exchange-biased $M$-$H_z$ hysteresis for the $t_{Cr_2O_3}$ = 7 nm sample.

**Structural characterization**

We grew Cr$_2$O$_3$ films in the thickness range from 5 nm to 20 nm on Y$_3$Sc$_2$Ga$_3$O$_{12}$ (YSGG) (111) substrates by using the pulsed laser deposition method. A 3 nm Pt layer was then deposited on top by using magnetron sputtering (Methods). $M$-$H$ loops show typical field-dependent magnetization of an AFM at different temperatures ($T$) (Supplement I). Also visible is a tiny kink around zero field due to the existence of uncompensated moments at the Cr$_2$O$_3$/Pt interface. Such uncompensated moments at the surface or interface in Cr$_2$O$_3$ systems have been successfully utilized to probe antiferromagnetic domain walls[30], flexomagnetism and vertically graded Néel temperatures[31]. Fig.1**b** illustrates a cross-sectional image of YSGG//Cr$_2$O$_3$/Pt obtained via scanning transmission electron microscopy (STEM). The image with atomic resolution confirms the epitaxial growth of Cr$_2$O$_3$ on the YSGG substrates. Its Fourier-transformed pattern reveals that the $c$ axis of Cr$_2$O$_3$ is out-of-plane (Fig. S2a). The X-ray magnetic linear dichroism (XMLD) spectrum of the Cr L-edge is used to characterize the Néel vector of Cr$_2$O$_3$. Fig.1**c** presents the data for the 7 nm Cr$_2$O$_3$ structure at 180 K. The XAS spectrum is typical for Cr$_2$O$_3$, representing exclusively the expected Cr$^{3+}$ ions. The experimental geometry is schematically depicted in the inset of Fig.1**c**. The XMLD spectrum is then obtained as XMLD=XAS$_{//}$−XAS$_{\perp}$ (Methods), which exhibits a zero-negative-positive-zero oscillating feature around $L_3$ and $L_2$ edges, which is characteristic of a Néel vector along the out-of-plane direction[32].

Moreover, the exchange bias effect associated with the exchange coupling between FM and AFM films at their interface constitutes an additional evidence of AFM order[33]. We characterized the exchange bias field in the YSGG//YIG(5 nm)/Cr$_2$O$_3$($t_{Cr_2O_3}$)/Pt(3 nm) control samples. The exchange bias field increases with enhancement of the AFM ordering and relates to the Néel temperature ($T_N$) of the $Cr_2O_3$ film. We chose to grow YIG films with PMA as an exchange-biased FM that also grows epitaxially on YSGG and hence did not change the atomic structures of $Cr_2O_3$ on top as confirmed by STEM (Supplement II). Fig.1d shows the exchange bias field $H_{eb} \equiv (H_{c+}+H_{c-})/2$ of YSGG//YIG/$Cr_2O_3$ for various values of the $Cr_2O_3$ thickness ($t_{Cr_2O_3}$). The negative $H_{eb}$ of all samples is non-zero below the blocking temperature $T_B$ for $t_{Cr_2O_3}$=7 nm to 20 nm. The inset figure shows a representative $M$-$H_z$ hysteresis at 150 K. These data indicate that the AFM order is stabilized below 300 K in all samples, as expected based on the bulk compound.

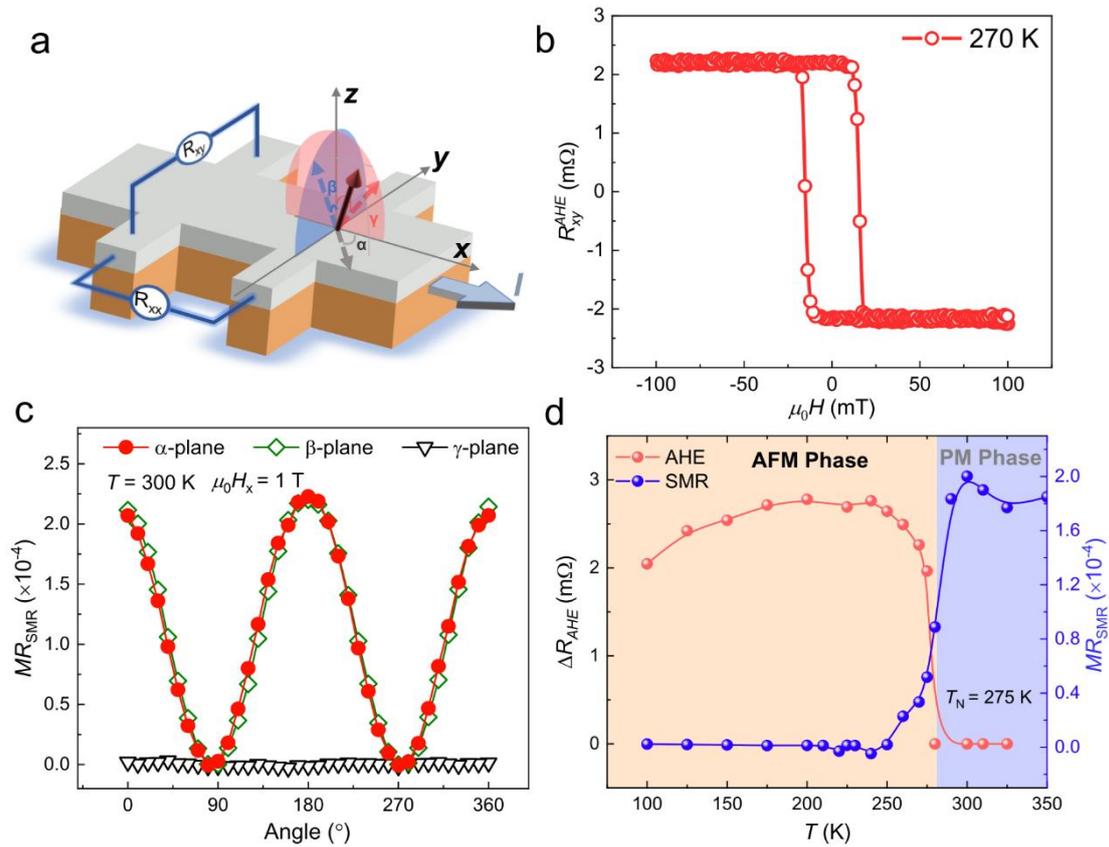

**Figure 2. AFM ordering of $Cr_2O_3$ as a function of temperature. a**, The experimental geometry to measure the AHE and SMR effects. The definition of the three angles $α$, $β$ and $γ$ are indicated and they rotate in the xy, yz and xz plane respectively. **b**, The dependence of the anomalous Hall resistance $R_{xy}$ on the out-of-plane field $H_z$ for the YSGG//$Cr_2O_3$(7)/Pt(3 nm) device. **c**, The angular-dependences of the longitudinal

resistance change ($\Delta R_{xx}$) for the $\alpha$, $\beta$ and $\gamma$ scans at 300 K under $\mu_0 H_x$=1 T. **d**, The $T$-dependence of the saturated $\Delta R_{xy}$ and the SMR ratio, $\Delta R_{AHE}$ approaches zero above 275 K.

## AHE and $T$-dependence of AFM ordering

We further characterized the AFM behavior of $Cr_2O_3$ by magneto-transport measurements (Fig.2). The stack with $t_{Pt}$=3 nm was patterned into Hall bars as shown in Fig.2**a**. We measured the anomalous Hall resistance ($R_{xy}$) as a function of the out-of-plane magnetic field $H_z$ to quantitatively estimate the AHE induced by the uncompensated spins at the $Cr_2O_3$/Pt interface at various temperatures. Fig.2**b** shows a clear hysteresis in the $H_z$-dependence of $R_{xy}$ for $t_{Cr_2O_3}$ = 7 nm at 270 K. Fig.2**d** summarizes the $T$-dependence of the remanent resistance $\Delta R_{AHE} \equiv [R_{xy}^{up} - R_{xy}^{down}]/2$ at $\mu_0 H$=0 mT: $\Delta R_{xy} \approx 0$ as $T \geq 275$ K, indicating a paramagnetic (PM) phase of $Cr_2O_3$ while $\Delta R_{xy}$ increases abruptly once $T <$ 275 K and a hysteresis in Fig.2**b** emerges, suggesting the formation of a magnetically ordered phase. This $T$-dependence of the AHE resistance is well described by a recent independent theoretical study in which a perpendicular surface spin state of $Cr_2O_3$ emerges below $T_N$[34]. As $T$ decreases further, $H_c$ increases and eventually saturates, indicating an enhanced AFM order and thus a PMA field ($H_K$) (Supplement III). Interestingly, the AHE observed here is different from Ref.[35] where the AHE was attributed to a nonzero parasitic magnetization due to doping. Here, in contrast, using polarized neutron reflectometry (PNR), we confirm no parasitic magnetization inside our films (Supplement IV).

Furthermore, we also measured the dependence of the longitudinal resistance ($\Delta R_{xx}$) on the direction of the applied field of 1 T at different $T$. Here, current is applied along the x axis and the spin current induced by the SHE is then polarized in the y axis. Fig.2**c** shows the angular-dependence of $\Delta R_{xx}$ at 300 K in the PM phase. $\Delta R_{xx}$ = 0 for the $\gamma$ scan when the spin polarization is always perpendicular to the spins in $Cr_2O_3$ but oscillates in a 180° period for the $\alpha$ and $\beta$ scans when the spins in $Cr_2O_3$ are driven alternatively collinear or perpendicular to the spin polarization. This observation aligns well with those of the spin Hall magnetoresistance (SMR) picture, thereby ruling out an anisotropic magnetoresistance (AMR) origin and the magnetic proximity effect in Pt. The AHE can be explained by the SMR origin and a complex spin-mixing conductance at the Pt/$Cr_2O_3$ interface[36]. Fig.2d presents the $T$-dependence of the SMR ratio ($\Delta R_{xx}/R_{xx}$). The SMR ratio of the $Cr_2O_3$/Pt bilayers decreases sharply as $T$ approaches $T_N$ ~275 K and tends to zero as $T$ is decreased well below $T_N$ ($T \ll T_N$) (Fig.2d). Once the AFM order is firmly established, the applied field of 1 T is overwhelmed by $H_K$ and cannot arbitrarily rotate the spins in $Cr_2O_3$, so that the SMR effect is quenched in the AFM phase and only arises in the PM phase as reported in $Al_2O_3$//$Cr_2O_3$/Pt bilayers[37]. In contrast, $R_{xy}$ increases steeply around $T_N$ when decreasing $T$. As $T > T_N$, there are no ordered interfacial spins and $R_{xy}$ is negligibly small. Once $T \leq T_N$, the growing interfacial spins quickly give rise to a noticeable $R_{xy}$ which enables us to effectively read out the magnetic state of $Cr_2O_3$. Both SMR and AHE results demonstrate an AFM-PM transition around 275 K for the 7 nm $Cr_2O_3$ films, consistent with the exchange bias data in Fig.1d. Specifically, we confirm that the AHE induced by uncompensated interfacial spins is strongly related to the bulk AFM order through the temperature-dependent measurements: the AHE vanishes above $T_N$ as determined by the SMR effect in Fig.2d and the exchange-bias effect in Fig.1d.

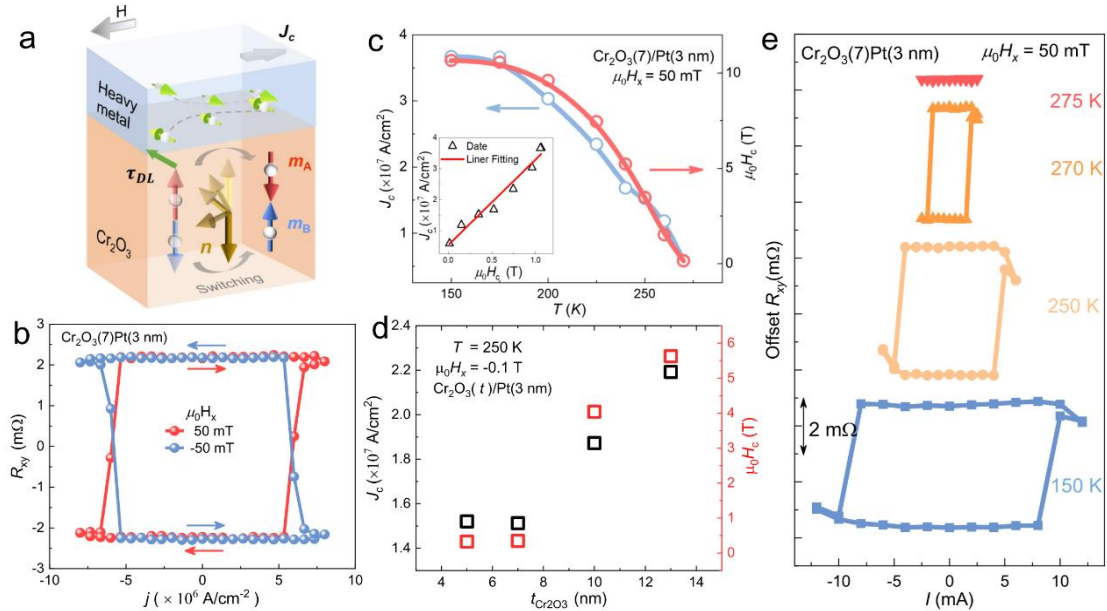

**Figure 3. SOT-induced switching of AFM order of the $Cr_2O_3$/Pt bilayers. A**, Schematics for switching the Néel order of $Cr_2O_3$ by SOT. **B**, The switching

performance of the $Cr_2O_3$/Pt device under opposite bias fields of $\pm 50$ mT. **c**, The $T$-dependence of $H_c$ and $J_c$ of the $Cr_2O_3$/Pt device. **D**, The thickness-dependence of $H_c$ and $J_c$ of the $Cr_2O_3$/Pt device. **E**, The switching loops at different $T$.

**Perpendicular and 180º switching of collinear AFM order by SOT**

Next, we demonstrate the 180º switching of the Néel order induced by SOT in $Cr_2O_3$/Pt heterostructures. In the top Pt layer, a charge current generates a SHE-induced spin current that is subsequently absorbed by uncompensated spins in $Cr_2O_3$, exerting an anti-damping-like torque on them and switching their directions. Finally, the switching of the uncompensated spins brings about the switching of the Néel order of the whole $Cr_2O_3$ (Fig.3**a**). Fig.3**b** shows the current-induced Néel order switching loops of $Cr_2O_3$ with an in-plane field $\mu_0 H_x = \pm 50$ mT at 270 K. Significantly, we find $\Delta R_{xy}$ at $I=0$ mA reaches 100% of the value measured in the $R_{xy}$-$H_z$ curve, indicating complete switching. Moreover, the switching direction, clockwise or counterclockwise, is controlled by the sign of $H_x$, which is a discriminative feature of the z-type SOT switching scheme[25]. When we replaced Pt by W, with an opposite spin Hall angle, the switching direction was also reversed (Supplement V). We also replaced Pt by Cu, and no switching loop was observed (Supplement V). Furthermore, we calibrated the real temperature $T^*$ of our sample with the current pulse turned on. The real sample temperature $T^*$ was still below $T_N$ (Supplement V). These features provide clear evidence of (1) 180º switching of the Néel order in $Cr_2O_3$ driven by SOT and (2) the dominant role of the SOT mechanism rather than the Oersted or heating mechanisms.

Note that the critical current ($J_c$), $5.8 \times 10^6$ A/cm$^2$ at 270 K, is lower than those reported in other nonmagnetic metal (NM)/AFM insulator systems ($J_c \approx 3 \times 10^7$ A cm$^{-2}$)[20,38] (CoO or $Fe_2O_3$) and comparable to those reported in the NM/AFM metallic systems[24]. The smaller $J_c$ is due to the weakened uniaxial anisotropy of $Cr_2O_3$ as $T$ approaches $T_N$, as illustrated by the strong $T$-dependence of the switching loops between 150 K and 275 K. The bipolar switching loops occur down to 150 K but vanishes at 275 K (Fig.3**e**). Notably, $J_c$ increases commensurately with $H_c$ as $T$ is decreased, suggesting a strong correlation with anisotropy. To test how robust the magnetic properties of collinear

AFM films are against external perturbations, we have investigated the SOT switching behavior under increased $H_x$ at lower temperature (Supplementary Fig. S7). SOT switching is even achieved under $\mu_0 H_x$=3 T at 150 K, demonstrating the robustness of this switching behavior against an external field (Supplement VI). This switching feature can support robust binary writing, making AFM $Cr_2O_3$ a viable memory candidate.

To quantitatively demonstrate the role played by the uniaxial anisotropy ($H_K$) of $Cr_2O_3$, we have systematically studied the dependence of $J_c$ on $H_x$, $T$ and $t_{Cr_2O_3}$, finding that $H_K$, $H_c$, and $J_c$ all increase as $T$ is reduced (Fig.3**c**). We also measured the switching loops for YSGG//$Cr_2O_3$($t_{Cr_2O_3}$)/Pt(3 nm) with $t_{Cr_2O_3}$ = 5 nm - 20 nm (Supplement VII). The thicker $t_{Cr_2O_3}$ results in the higher $H_c$, especially for thicker samples ($t_{Cr_2O_3}$ = 13 nm) with up to 5.5 T coercivity. The further increase in $t_{Cr_2O_3}$ leads to higher $H_c$ beyond the limit of our equipment, so we did not obtain their $R_{xy}$-$H_z$ hysteresis loops for the thicker samples ($t_{Cr_2O_3}$ >13 nm) below $T_N$. Correspondingly, $J_c$ also increases with increasing $t_{Cr_2O_3}$ for the samples with $t_{Cr_2O_3}$ ≤ 13 nm (Fig.3**d**), having a linear dependence on $H_c$ or $H_K$. While this linear dependence is similar to the z-type mode observed in HM/FM bilayers, a key difference from the HM/FM systems is that $J_c$ is insensitive to $H_x$. These features are well reproduced by a theoretical model in which a much stronger AFM coupling than anisotropy is supposed (Supplement VIII a). This switching insensitivity to the external field is beneficial in achieving a strong robustness against external field disturbance for memory applications.

To further gain insight into the SOT switching results, we use a theoretical framework that combines a time-dependent nonequilibrium Green's-function (TDNEGF) algorithm with Landau-Lifshitz-Gilbert (LLG) equation to calculate the dynamics of magnetization switching[39,40] (Supplement VIII a). Through the s-d exchange coupling, the polarized electrons exert a torque on the localized magnetic moments at the Pt/AFM interface, which reverses the AFM Néel vector. The simulations also show that the critical current is positively correlated with the anisotropic energy and insensitively correlated with the applied magnetic field, which is in accordance with experiments. On the other hand, we also studied the significance of uncompensated magnetic moments on the surface during the Néel order reversal process (Supplement VIII b). The above simulations conclusively show that differences in anisotropy, interlayer coupling, and especially magnetization of the interfacial region from that of the bulk $Cr_2O_3$ plays a critical role for the presence of the field-induced and SOT-induced Néel order switching.

**SOT-induced switching of the YIG/$Cr_2O_3$/Pt trilayer**

Inspired by the switching of $Cr_2O_3$, we expect SOT can drive perpendicular magnetization switching of YIG for YSGG//YIG/$Cr_2O_3$/Pt samples, since an interfacial coupling between YIG and $Cr_2O_3$ is verified in Fig.1**d**. For this purpose, we first measured $R_{xy}$ of the trilayer at different $T$. A sizable $R_{xy}$ below $T_N \approx 275$ K confirms the presence of the perpendicular AFM ordering (Supplement IX). Next, we measured the $R_{xy}$-$I$ loops to ensure that the Néel order of $Cr_2O_3$ in the trilayer sample can also be reversed by current. A binary switching hysteresis is also clearly observed with the reversible directions controlled by $H_x$, consistent with the bilayer devices, which indicates that SOT also acts effectively on the YIG/$Cr_2O_3$ system (Fig.4**d**). However, the magnetization state of YIG cannot be probed by the AHE since Pt is only connected with $Cr_2O_3$. Thus, we further conducted the in-situ current-induced switching measurement via the magneto-optical Kerr effect (MOKE) for the trilayer.

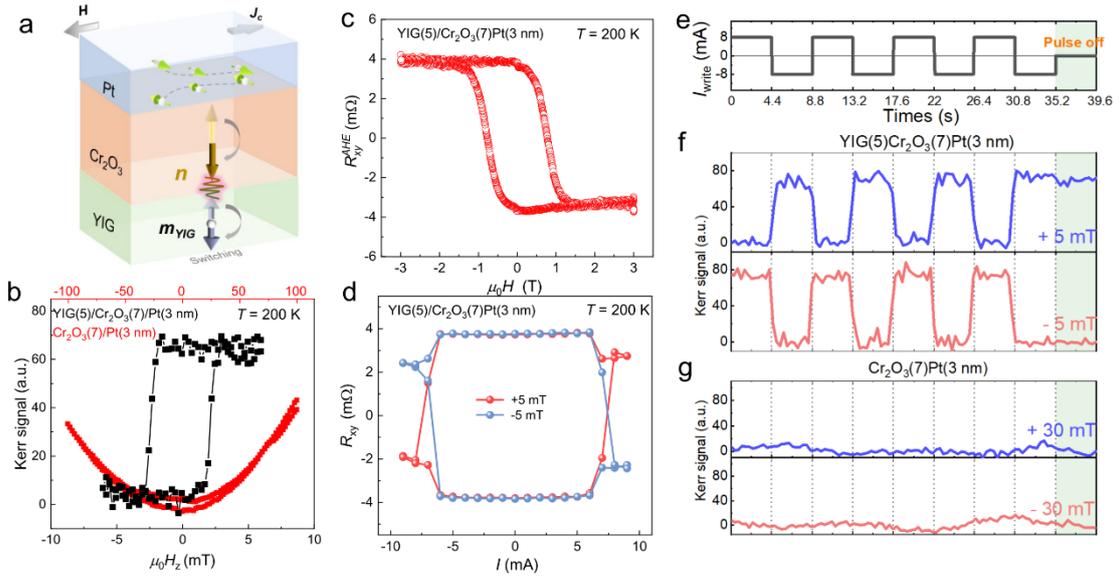

**Figure 4. SOT-induced switching of the YIG/$Cr_2O_3$/Pt trilayer. a**, The AFM-mediated switching schematics for the YIG/$Cr_2O_3$/Pt trilayer. **b**, The magnetization loop

detected by MOKE for the YIG/$Cr_2O_3$/Pt trilayer. No hysteretic MOKE signal was obtained for the $Cr_2O_3$/Pt bilayer. **c**, The $H_z$-dependence of the anomalous Hall resistance of the trilayer. **d**, The current-induced $R_{xy}$ switching for the trilayer. The switching direction depends on the current-collinear field. **e**, The current pulse sequence. **f**, The observed states by MOKE after each current pulse in **e**. The switching of the YIG magnetization was clearly identified with reversed switching directions at ±5 mT. **g**, No switching signal was detected for the $Cr_2O_3$/Pt bilayer.

Thanks to the negligible magnetization and transparency of $Cr_2O_3$, the Kerr signal can be safely attributed to the magnetization reversal of YIG (Fig.4**b**), allowing us to directly observe the YIG magnetization in the switching experiment. We successively applied positive and negative current pulses into the Hall bar channel (Fig.4**e**). The observed MOKE signal for the YIG/$Cr_2O_3$/Pt trilayer was recorded *in-situ* after applying the current pulses and shown in Fig.4**f**. Clearly, the high and low Kerr signals closely depending on current polarity correspond to the spin-up and spin-down states of YIG in Fig.4**e**&**f**. Furthermore, the magnitude of the current-induced Kerr signal change in Fig.4**f** is in excellent agreement with that of the field-induced YIG layer switching shown in Fig.4**b**, indicating full switching of the YIG magnetization. Both magnetization states of YIG can be reproducibly written. To further verify that the alternating Kerr signals originate in SOT-induced switching of the YIG magnetization, we changed the $H_x$ direction. As a result, the polarity of the MOKE signal is also reversed (Fig.4**f**). When the current is switched off, the final state is maintained, confirming its nonvolatility (the green region in Fig.4**f**). In the control experiment, we also sourced the alternating current pulses to the former $Cr_2O_3$/Pt bilayer sample. No Kerr signal switching is observed (Fig.4**g**). The magnetization switching of the YIG layer without direct connection with Pt unambiguously demonstrates the switching of the entire Néel vector of $Cr_2O_3$, as it cannot be explained merely by the switching of the interfacial uncompensated spins at the $Cr_2O_3$/Pt interface.

We note that several AFM-based magnon-transfer torque (MTT) effects have recently been reported to enable FM magnetization switching in FM/AFM insulator (AFI)/HM systems [41,42]. In this scenario, spin accumulation at the HM/AFI interfaces generates or annihilates magnons inside AFI and thus drives a magnon current across the AFI to deliver spin-angular momenta and eventually reverse the magnetization of FM by the spin-torque transferred at the FM/AFI interface. In this process, the spin polarization must be collinear with the Néel vector of the AFI, as is required for magnon generation or annihilation[43-44]. In stark contrast, here, the spin polarization is perpendicular to the Néel vector of $Cr_2O_3$, making the MTT effect impossible. Thus, we propose that the magnetization switching of YIG is mediated by the switching of the $Cr_2O_3$ Néel vector rather than a magnon current.

To experimentally rule out the MTT mechanism, we measured the SMR of a YIG/$Cr_2O_3$/Pt trilayer. If the MTT were to dominate, the YIG layer would have absorbed magnon current at the YIG/$Cr_2O_3$ interface so that the YIG magnetization should have influenced the Pt resistance according to the SMR mechanism. However, no SMR is observed below $T_N$ for the trilayer (Supplement X), indicating the spin current produced in Pt has already been fully absorbed by $Cr_2O_3$ before reaching the YIG layer. The $T$-dependence of SMR is consistent with the experimental data in Ref.[45]. These observations strongly favor a picture in which transferred spin torque from Pt is directly imposed on the perpendicular Néel vector of $Cr_2O_3$, resulting in 180º switching of its Néel order and indirectly reversing the YIG magnetization through exchange coupling with $Cr_2O_3$ (Fig.4**a**).

Thus, our experiment reveals AFM-mediated torque as a new pathway for energy-efficient magnetization switching of magnetic insulators. This process is as energy efficient as MTT. Such a picture is also well reproduced by theoretical calculation, which reveal that the $Cr_2O_3$ Néel vector and YIG magnetization are simultaneously switched by a polarized spin current (Supplement VIII a). In addition, the AFM-mediated torque mechanism which can nonlocally switch magnetization of a magnetic insulator embedded in AFMI/FM heterostructures can be potentially used to electrically control magnetic states of emerging insulating magnon devices, such as magnon junctions[46,47].

From the above observations and analysis, it is clear that this magnetic insulator / AFM insulator / heavy metal trilayer system with perpendicular anisotropy is required to unambiguously demonstrate the 180º SOT switching of the AFM Néel order. Encouragingly, after proving the SOT-switching feasibility in this collinear and insulating AFM system, we see no physical limitation to prevent the same mechanism from being generalized to other collinear AFM systems such as metallic MnN[48], FeRh[49], $Mn_5Si_3$, CrSb, $RuO_2$[13] and so on. Some of them, for example $RuO_2$, has been predicted potentially to develop AFMTJs with ultrahigh TMR ratios of above 500% [13].

The performance of electrical manipulation of the antiferromagnetic order is crucial for practical memory devices. More specifically, the spin current generated by sub-terahertz antiferromagnetic resonance has been reported in $Cr_2O_3$, offering attractive dynamic properties for potential applications in ultrafast devices. Accordingly, we also switched the Néel vector of $Cr_2O_3$ by 6-ns current pulses, a record-fast value for SOT-switching of an AFM insulator system, which is also a key step toward ultrafast and robust device applications[7] (Supplement XI). Further, the SOT switching endurance performance was tested through the application of 5000 alternating pulse currents. The obtained switching curves did not show noticeable degradation, demonstrating high endurance (Supplement XII). We note that $T_N$ of $Cr_2O_3$ is 275 K currently, insufficient for room-temperature applications. To mitigate this shortcoming, epitaxial or chemical strain engineering has been recently demonstrated as a viable path to increase $T_N$ to 400 K,[31,50,51] which renders $Cr_2O_3$ relevant for industrial microelectronics applications.

**Conclusions**

In summary, our study unambiguously demonstrates efficient control over the Néel order of collinear AFM insulator $Cr_2O_3$ with perpendicular magnetic anisotropy by spin-orbit torques. The AFM order is switched with a minimum current density of ~5.8×10$^6$ A/cm$^2$. Furthermore, the electrical manipulation exhibits excellent endurance and high-speed switching performances. With both efficient SOT-writing and reliable

AHE reading, the collinear and perpendicular AFM $Cr_2O_3$ can be a promising candidate for an AFM memory with fully electrical operation capability, paving the road toward the long-sought AFM spintronics with low-energy consumption, high-density, ultra-fast speed and strong robustness against disturbance.

**Methods**:

**Sample and device fabrication**

The high quality $Cr_2O_3$ layers with different thicknesses were prepared by pulsed laser deposition with a KrF excimer laser. The YSGG substrate temperature was kept at 723 K and $O_2$ pressure of 20 mTorr (1 Torr = 133.3 Pa) during deposition. The deposition rate of $Cr_2O_3$ was around 0.5 nm min$^{-1}$. Then the samples were transported into magnetron sputtering to deposit Pt or W layer (3 nm) on $Cr_2O_3$ at room temperature. For YIG/$Cr_2O_3$/Pt trilayer samples, we first grew a 5 nm thick single-crystalline YIG on (111)-oriented YSGG substrates using magnetron sputtering and annealed at 800 °C for an hour. Subsequently, $Cr_2O_3$ and Pt were grew as described above.

Structural analysis was performed by the high-resolution transmission electron microscope (HRTEM). The cross-sectional TEM image of $Cr_2O_3$/Pt and YIG/$Cr_2O_3$/Pt samples revealed high quality $Cr_2O_3$ and YIG layer, which laid the foundation for us to do the transport and switching measurements. What's more, the Fourier transformation of the HRTEM images of $Cr_2O_3$/Pt and YIG/$Cr_2O_3$/Pt samples showed the same epitaxial relationship, that is, the (111)-oriented YSGG(YIG) epitaxial the (0006)-oriented $Cr_2O_3$.

Magnetization measurements were carried out in a commercial superconducting quantum interference device magnetometer (MPMS3, Quantum Design). To obtain the signal from only the multilayers, the diamagnetic contribution of the YSGG substrate is separately measured and subtracted from the total M. The $Cr_2O_3$ magnetization loops are shown as a function of out-of-plane magnetic field within ±4 T at 100 K and 300 K in supplement. As for YIG/$Cr_2O_3$/Pt trilayer samples, we first carried out a field cooling process and then measured M-H loops to obtain the exchange bias field at different temperatures. The exchange bias field $H_{eb}$ of YSGG//YIG/$Cr_2O_3$ for various $t_{Cr2O3}$ is shown in Figure 1d. The XAS and XMLD measurements were performed at Vector magnetic system in one of the SSRF BL07U end stations for sample YSGG//Cr$_2$O$_3$(7)/Pt(1 nm) at 200 K which is below the T$_N$. The polarization dependence was obtained with linearly polarized light with an incident angle of 45° from the c-axis, which yields near $E_\perp$ or $E_\parallel$ in a fixed experimental geometry by utilizing the planar or vertical polarization selectivity of the elliptically polarized undulator. The XMLD spectra were obtained by subtracting $E_\perp$ from the $E_\parallel$ spectra. For the transport and switching measurements, the multilayers are patterned into Hall bar devices with in-plane dimensions of 10 μm × 100 μm using standard photolithography combined with an Ar$^+$ etching process, and Pt/Au contact pads and wires are deposited by magnetron sputtering.

**PNR measurements**

PNR measurements were performed using the POLREF instrument at the ISIS Neutron and Muon Source. The POLREF beamline is a white beam Time of Flight (ToF) polarized neutron reflectometer with a polarized wavelength band of 0.2 nm - 1.4 nm. The samples were mounted horizontally in a helium flow cryostat within an electromagnet with a maximum field of ±700 mT that can be applied in the plane of the sample. A helium exchange gas pressure of 4 kPa was used, providing a temperature stability of ±0.01K. The samples were held in place by gravity alone to avoid the use of epoxy, which may strain the film and substrates upon cooling to lower temperatures. The data were reduced using the Mantid framework in combination with additional in-house python scripts and analyzed using the Refl1D software package. Uncertainties on fitted parameters were obtained using the DREAM algorithm in the Bumps software package, a Markov-chain Monte Carlo (MCMC) method.

**Transport measurements**

The transport characteristics were measured by a standard four-probe method using a physical property measurement system (PPMS, Quantum Design). First, we measured the transverse resistance $R_{xy}$ as a function of the $H_z$ to obtain anomalous Hall resistance. After subtracting a linear background due to the ordinary Hall effect, a clear hysteresis was observed when *T* was far below 275 K. As for the SMR measurement, we measured longitudinal resistance $R_{xx}$ with the rotation of devices in the $\alpha, \beta, \gamma$ plane. For the switching measurement, we applied a 100 ms pulse write current $I_{write}$ to switch antiferromagnetic domains. Then a d.c. read current of $I_{read}$ = 0.1 mA was applied along the x direction, where a wait time of 500 ms was inserted after turning off the write current and before measuring $R_H$.

**MOKE measurements**

The YIG layer switching experiments were measured by the Evico magneto-optic Kerr microscope (MOKE), which facilitates the visualization of magnetization switching processes as well as optically recording magnetization curves on magnetic materials. MOKE system is also equipped with external electrical measurement system including one Keithley 6221 source meters and a Keithley 2182 source meter used to provide currents and measure the Hall voltage, respectively. We first measured the magnetization curves of samples YSGG//$Cr_2O_3$(7)/Pt(3 nm) and YSGG//YIG(5)/$Cr_2O_3$(7)/Pt(3 nm) dependent on the out-of-plane magnetic field at 200 K. The obvious magnetization curves were observed in trilayer samples but vanished in bilayer samples, manifesting the PMA of YIG and negligible magnetization of $Cr_2O_3$. It should be noted that no clear change of MOKE image is observed from MOKE microscopy YIG magnetization switching, which may be caused by the thinner YIG thickness in combination with the weak magnetization. To investigate the magnetization dynamics of YIG layer during SOT switching, a time-resolved-MOKE model is used. For the current switching experiment, a Keithley 2612B source meter was used to apply current pulses with a pulse length of 1 ms through the devices, and the magnetization states were captured by Kerr signal using Kerr microscope at the same time. The measurement sequence with different in-plane $H_x$ values is shown in Fig. 4.

**Data Availability**: The datas that support the findings of this study are available from the corresponding author upon reasonable request.

**Code availability**: The codes that support the theoretical modelling of this study are available from the corresponding authors upon reasonable request.

**Acknowledgements**: The XMLD measurements were carried out at BL07U beamline of Shanghai Synchrotron Radiation Facility (SSRF). The authors appreciate the financial support from the National Key Research and Development Program of China (MOST) (Grant No. 2022YFA1402800, 2022YFA1402603), the National Natural Science Foundation of China (NSFC) [Grant No. 51831012, 12134017,

11974398, 52161160334, the Strategic Priority Research Program (B) of Chinese Academy of Sciences (CAS) [Grant No. XDB33000000. C.H.W appreciates financial support from Youth Innovation Promotion Association, CAS (2020008). We thank the ISIS neutron and muon source for the provision of neutron beamtime (RB2220260, https://doi.org/10.5286/ISIS.E.RB2220260). Certain commercial equipment, instruments, software, or materials are identified in this paper in order to specify the experimental procedure adequately. Such identifications are not intended to imply recommendation or endorsement by NIST, nor it is intended to imply that the materials or equipment identified are necessarily the best available for the purpose.

**Contributions**: X.F.H led and was involved in all aspects of the project. W.Q.H and Y.J.Z, and C.S deposited stacks and fabricated devices. W.Q.H and C.H.W conducted magnetic and transport property measurement. W.Q.H performed the MOKE measurements with assistance from B.S.C and T.Y.G. T.Y.Z, C.H.W and W.Q.H contributed to modelling and theoretical analysis. W.Q.H, J.H.X conducted the XAS and XMLD experiment in SSRF. A. J. G, P. P. B, C. K., A. C., S. L., and H.W. conducted the polarized neutron scattering experiments at the ISIS facility. W.Q.H, C.H.W and X.F.H. wrote the paper. X.F.H and C.H.W supervised and designed the experiments. All the authors contributed to data analysis.

**Conflict of Interests**: The authors declare no competing interests.